\documentclass[12pt]{article}
\usepackage{graphicx}
\begin{document}

\title{Infrared afterglow of  GRB041219 as a result of reradiation on
dust in a circumstellar cloud.}

\author{Barkov M.V., Bisnovatyi-Kogan G.S.\thanks{Space Research Institute,
Moscow. Emails: barmv@iki.rssi.ru, gkogan@iki.rssi.ru}}

\date{\today}

\maketitle

\begin{abstract}
Observations of gamma ray bursts (GRB) afterglows in different
spectral bands provide a most valuable information about their
nature, as well as about properties of surrounding medium.
Powerful infrared afterglow was observed from the strong
GRB041219. Here we explain the observed IR afterglow in the model
of a dust reradiation of the main GRB signal in the envelope
surrounding the GRB source. In this model we do not expect
appearance of the prompt optical emission which should be absorbed
in the dust envelope. We estimate the collimation angle of the
gamma ray emission, and obtain restrictions on the redshift
(distance to GRB source), by fitting the model parameters to the
observational data.
\end{abstract}

\section{Observational data}

Observations of gamma ray bursts (GRB) afterglows in different
spectral bands provide a most valuable information about their
nature, as well as about properties of surrounding medium.
Powerful infrared afterglow was observed from the strong
GRB041219, recorded by  INTEGRAL and SWIFT  \cite{2866,2874}. By
lucky chance IR observations started as soon as 2.4 minutes after
GRB registration, during the GRB itself (second such case after
GRB990123), which lasted 520 seconds, and was very strong
$F_{\gamma}\sim 10^{-4}$ erg/cm$^2$. The IR flux (K - band)
corresponded to $K = 15.5^m$ 2.4 minutes after registration
\cite{2872}, $K=14.9^m$ after 0.8 hour, $K = 15.5^m$ after 1.55
hours \cite{2876}, and $K=16.5^m$ after 1.01 day \cite{2884}.
Observations 47.25 hours after the burst had shown $K=17.6^m$,
$H=18.9^m$, and $J=19.9^m$ \cite{2916}. No optical afterglow was
registered 74 seconds after registration up to $17.2^m$
(unfiltered) \cite{2868}, and the limiting value $R=19.4^m$ was
obtained in \cite{2889} for the prompt optical emission of this
burst. A weak growing radio emission at 4.9 GHz was detected from
this GRB, with fluxes 205 microJy 1.75 days after the burst
\cite{2894}, and 349 microJy 2.74 days after GRB \cite{2895}.

\section{Model}

The plot for the observed IR flux on the time $L_{IR}(t)$ is
represented in Fig.1. The IR flux in $K=14.9^m$ is equivalent to
$2.4\times 10^{-13}$ erg/(s $\cdot$ cm$^2)$. Accepting the
duration of IR emission during 1 day we obtain the IR fluence
$\sim 10^{-8}$ erg/cm$^2$, what corresponds to $10^{-4}$ of the
total energy of GRB. With account of the limited spectral range of
SWIFT data, we expect the total fluence to be $F_{\gamma} \approx
2 \cdot 10^{-4}$ erg/cm$^2$. Suppose that the observed $IR$
emission is formed due to interaction of the GRB main pulse with
the surrounding dust envelope. The mechanism of this formation is
the following. Gamma photons with energy exceeding 10 keV interact
with the matter, including dust, mainly due to Compton scattering.
The photons with energies $E_{\gamma}=10-100$ keV interact with
the electron inside the dust grain, transfering $\xi_e$ of its
energy to this electron. The electron is absorbed inside this
grain, the dust grain reradiate this energy in IR, and gamma ray
photon is flowing away. We consider only sufficiently distant
regions of the dust envelope, where the dust evaporation is not
important. IR reradiation is produced effectively only in such
condition \cite{boch,burke,drain}.

\begin{figure}[tbp]
 \includegraphics[width=5in]{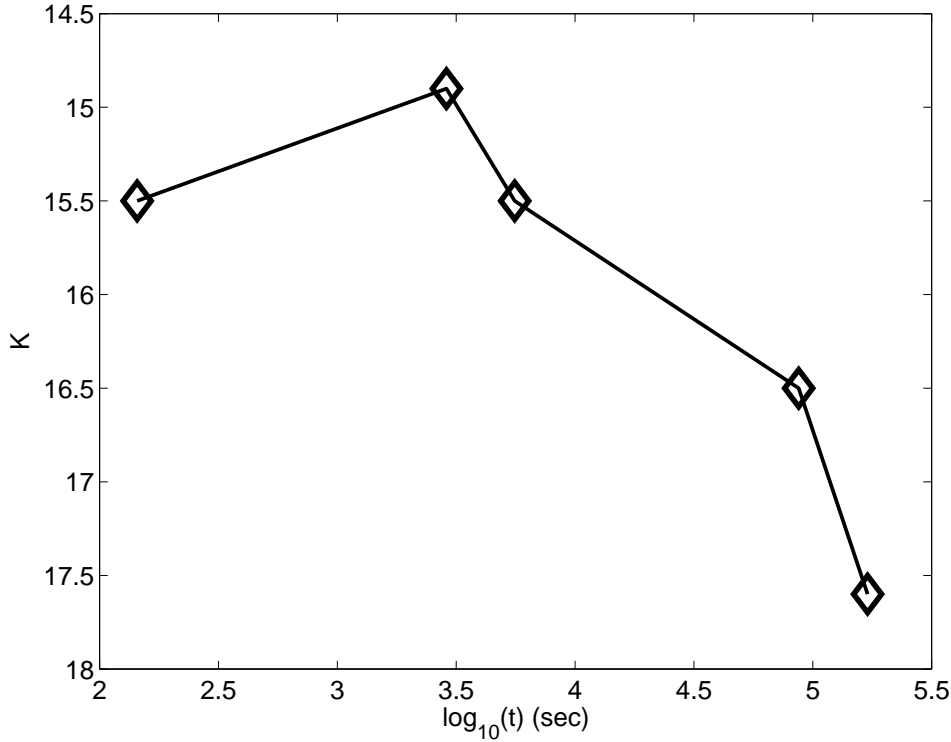}
 \caption{ Light curve of GRB041219 in band K, time in seconds.}
 \label{fig2}
\end{figure}
The dust formed by heavy elements contains  about 1\% of the whole
mass of the gas for solar abundances. XMM observations of
GRB011211 and  GRB030227 had indicated \cite{obr}, that abundances
of light elements (Mg, Si, S) in the $X$ - radiating matter more
than 10 times exceed the Solar ones. Taking 10\% for the mass of
the dust component, the amount of energy absorbed in the
surrounding envelope due to Compton scattering would be 10 times
larger than that absorbed by the dust. As a result the IR emission
should be equal to $\alpha_D \sim 0.1$ of the total energy
absorbed in the surrounding cloud, $E_{abs}\sim 10^{-7}$
erg/cm$^2$. The total absorbtion optical depth of the envelope
would be equal to $\tau_{abs} \sim 5 \cdot 10^{-4}$, if the gamma
ray and IR photons were radiated in the same body angle. For a
small gamma ray beam $\Theta \ll 4\pi$ and isotropic IR
reradiation the absorption of gamma ray emission should be
$4/\Theta^2$ times larger, and the total depth of the envelope to
the Compton interaction should be equal to

\begin{equation}
\label{tau}
 \tau_{c} \sim \frac{2 \cdot 10^{-3}}{\xi_e \Theta^2},\,\,\,
 \tau_{IR}=\alpha_D\tau_c \sim \frac{2 \cdot 10^{-4}}{\xi_e \Theta^2},
\end{equation}
where $\tau_{IR}$ is the optical depth of the dust component. The
surface number density $\Sigma_e$ for Thompson cross-section
$\sigma_T$, and $\xi_e=0.25$ for the characteristic GRB spectrum
\cite{barbk}, is equal to

\begin{equation}
\label{eq1} \Sigma_e=\frac{\tau_c}{\sigma_T} \sim
   \frac{3.2\cdot 10^{21}}{\Theta^2} {\rm cm}^{-2}.
\end{equation}
We suppose that GRB is a result of processes in the remnant of the
evolution of a massive star (i.g. heavy disc falling into a
massive black hole \cite{mFW}), which had lost a considerable part
$\ge 1/2$ of its matter which formed a massive cold envelope. The
massive star is embedded into the dense interstellar cloud. Take
for simplicity that the envelope is in the form of a thin
spherical shell with a radius $R$, and thickness $h\ll R$. We take
into account that the IR dust reradiation is very rapid, lasting
less than one second \cite{boch}, what is much less than all
characteristic times inherent to  GRB afterglow radiation.
Therefore the IR reradiation may be taken as an instant process.
The distant observer accept at given moment the emission
reradiated by the shell of the matter in the ring with the radius
$r \approx \sqrt{2cRt}$, here $t$ is counted from the time of GRB
detection, $r \ll R$, see Fig.2. The collimated gamma ray
radiation illuminates only part of the shell with the radius
$r_I=R\Theta$. The angular size $\theta$ of the ring from which
the radiation is coming to the observer, is equal to

\begin{equation}
\label{eq3} \theta \sim r/R = \sqrt{2ct/R}.
\end{equation}
The observer will see the IR reradiation until $\theta$ reaches
its maximal value $\Theta$, after with the IR  radiation abruptly
stopped in the thin shell model, but starts to drop more rapidly
in the realistic case. Consider for simplicity, that gamma ray
fluence, $\tau_c$, and $\tau_{IR}$ do not depend on the angle
$\theta$ inside the beam. Let us introduce the values of the total
energy of GRB $E_{\gamma}$ (erg), total energy of the infrared
afterglow $E_{IR}$ (erg), observed fluence of the GRB $F_{\gamma}$
(erg/cm$^2$), observed fluence of the infrared afterglow $F_{IR}$
(erg/cm$^2$), observed infrared flux $L_{IR}$ (erg/cm$^2$/s). Here

\begin{equation}
\label{flux}
  E_{IR}=\tau_{IR} E_{\gamma},\,\,\,
  F_{\gamma}=\frac{E_{\gamma}}{\pi\Theta^2 l^2},\,\,\,
  F_{IR}=\frac{E_{IR}}{4\pi l^2},
\end{equation}
where $l$ is the distance from the source to the observer. The
observed $IR$ flux $L_{IR}(t)$ of reradiation as a function of
time is determined as

\begin{equation}
\label{eq2} L_{IR}(t)=\frac{1}{4 \pi l^2} \frac{E_{\gamma}}{\pi
\Theta^2} \tau_{IR}
  2\pi \theta \frac{d\theta}{dt} =\frac{E_{\gamma} \tau_{IR}}{4 \pi l^2}
  \frac{2 c}{R \Theta^2}=\frac{1}{4 \pi l^2}\frac{E_{IR}}{t_{IR}},
\end{equation}
where $t_{IR} \sim 1$ day is the observed time of the $IR$
radiation over which the $IR$ flux may be taken almost constant,
and which is uniquely connected with the collimation angle
$\Theta=\sqrt{2ct_{IR}/R}$. As follows from (\ref{eq2}), the
observed $IR$ flux is constant in this simple model at $t<
t_{IR}=R\Theta^2/2c$, and becomes zero at larger $t$. For
nonuniform gamma ray fluence, and optical depth inside the beam
the IR flux is variable at $t< t_{IR}$, and does not become zero
at $t> t_{IR}$ for a thick dust shell.

\begin{figure}[tbp]
 \includegraphics[width=5in]{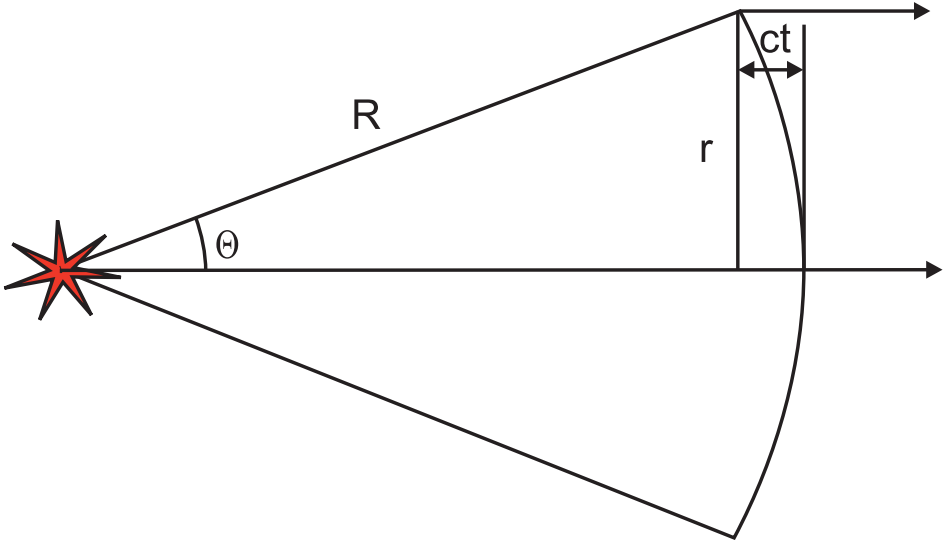}
 \caption{ The scheme of the model. The arrows are direction to distant observer,
 $r=\sqrt{2cRt}$.}
 \label{fig1}
\end{figure}

\section{Estimations of GRB041219 properties}

Let us do quantitative estimations of the model parameters for
GRB041219. Starting from the observed values $F_{\gamma}$,
$F_{IR}$, $t_{IR}$, and taking the optical depth as
$\tau_c=nR\sigma_T$, where $n$ is the average concentration in the
cloud, we obtain the expressions for $\Theta$, $R$, $n$ and
$\tau_c$ as follows

\begin{equation}
\label{par}
  n=\frac{\tau_c \Theta^2}{2c t_{IR} \sigma_T}
  \approx 2\cdot 10^6\, {\rm cm}^{-3},\,\,\,
  \Theta=\sqrt{\frac{2ct_{IR}}{R}}\approx
  \frac{0.077}{\sqrt{R_{18}}}=\frac{4.4^o}{\sqrt{R_{18}}},\,\,
  \tau_c=\frac{4R_{18}}{3},
\end{equation}
where $R_{18}=R/10^{18}$cm. Note, that estimations of the
collimation angle and properties of the media surrounding GRB
source do not depend on the distance to GRB, and therefore on its
energy production. The absence of the prompt optical emission
could be connected with a strong absorption inside our Galaxy,
because this GRB lies very close to the Galactic plane with
$b=0.6$ deg \cite{2916}, but we do not expect optical or soft
$X$-ray afterglows because of their absorption in the remaining
dust envelope. The afterglow $IR$ flux, connected with a secondary
reradiation of the $X$ ray and optical quanta by dust, will be
considerably fainter, than the direct $IR$ afterglow produced by
GRB itself. The matter heated by GRB gamma radiation up to
temperatures $10^6\,-\,10^7$ K has a cooling time of the order of
few weeks for the density (\ref{par}), what is about 100 times
larger than the duration of the prompt IR radiation accompanying
heating of the dust by GRB pulse \cite{bktim,barbk}. Therefore the
$IR$ flux at $t
> t_{IR}$, due to this reradiation should be about $3^m$
magnitude fainter than in the prompt IR source. In our model the
dust should not be evaporated by the GRB main pulse. According to
\cite{drain}. The critical gamma ray fluence which is destroying
the dust grain is

\begin{equation}
\label{lcrit}
  F_{\gamma,cr}=\frac{E_{\gamma,cr}}{\pi \Theta^2
\varepsilon_{\gamma} R^2}=4\times 10^{21}{\rm cm}^{-2},
\end{equation}
where $\varepsilon_{\gamma} \approx 100$ keV is the average energy
of the gamma ray quanta. Taking into account (\ref{par}), we
obtain the restriction in the GRB total power $E_{\gamma} \leq 1.3
\cdot 10^{49} R_{18}$ erg, and using the corrected observed gamma
ray fluence $F_{\gamma}=2\times 10^{-4}$ erg/cm$^{-2}$ and
(\ref{par}) we obtain the restriction to the distance $l$ and
redshift $z$ of this GRB

\begin{equation}
\label{lim}
  l \le 2\times 10^{27} R_{18}\,\, {\rm cm}=
  670\, R_{18}\,\,{\rm Mpc}, \quad
  z \le 0.16\, R_{18}.
\end{equation}
Taking into account, that Compton interaction should not decrease
considerably the GRB fluence, we put a restriction for $\tau_c
\leq 1$, corresponding to about of $\xi_e=0.25$ of the absorbed
gamma ray fluence \cite{bel_ill,barbk}, and obtain the following
restrictions to GRB041219 parameters

\begin{equation}
\label{lim1}
 R_{18} \leq 3/4, \,\,\,
 z \leq 0.12, \,\,\,
 E_{\gamma} \leq 10^{49}\, {\rm erg},\,\,\,
 \Theta \geq 5^o.
 \end{equation}
 Note that these restrictions directly depend on the estimations of dust
 evaporation condition (\ref{lcrit}).

\section*{Acknowledgements} We are grateful to N.G. Bochkarev for useful discussions.

\end{document}